\documentclass[aps,reprint,floatfix,superscriptaddress,prl]{revtex4-1}

\usepackage[usenames,dvipsnames]{xcolor}

\definecolor{icolor1}{HTML}{0000FF}
\definecolor{icolor2}{HTML}{FFA400}
\definecolor{icolor3}{HTML}{008B00}
\definecolor{icolor4}{HTML}{FF34A7}
\definecolor{icolor5}{HTML}{632280}
\definecolor{icolor6}{HTML}{5E3100}
\newcommand{\siv}{\textrm{SiV}}
\newcommand{\siva}{\textrm{SiV}_\mathrm{I}}
\newcommand{\sivb}{\textrm{SiV}_\mathrm{II}}
\newcommand{\nv}{\textrm{NV}}

\newcommand{\doubEg}{{}^2\mathrm{E}_\mathrm{g}}
\newcommand{\doubEu}{{}^2\mathrm{E}_\mathrm{u}}

\newcommand{\lambdag}{\lambda^\mathrm{SO}_\mathrm{g}}
\newcommand{\lambdau}{\lambda^\mathrm{SO}_\mathrm{u}}

\newcommand{\Gammaup}{\mathit{\Gamma}_\uparrow^\mathrm{ph}}
\newcommand{\Gammadown}{\mathit{\Gamma}_\downarrow^\mathrm{ph}}

\newcommand{\Ex}{\mathrm{E}_\mathrm{X}}
\newcommand{\Ey}{\mathrm{E}_\mathrm{Y}}

\newcommand{\Dthreed}{\mathrm{D}_\mathrm{3d}}
\newcommand{\transC}{\textrm{C}}

\usepackage{graphicx}
\newcommand*\autoref[1]{Figure \ref{#1}}

\begin{document}

\title{Indistinguishable photons from separated silicon-vacancy centers in diamond}

\author{A. Sipahigil}
\email{sipahigil@physics.harvard.edu }
\affiliation{Department of Physics,	Harvard University,	17 Oxford Street, Cambridge, MA 02138, USA}
\author{K. D. Jahnke}
\affiliation{Institute for Quantum Optics, University Ulm, Albert-Einstein-Allee 11, 89081 Ulm, Germany}
\author{L. J. Rogers}
\affiliation{Institute for Quantum Optics, University Ulm, Albert-Einstein-Allee 11, 89081 Ulm, Germany}
\author{T. Teraji}
\affiliation{National Institute for Materials Science, 1-1 Namiki, Tsukuba, Ibaraki 305-0044 Japan}
\author{J. Isoya}
\affiliation{Research Center for Knowledge Communities, University of Tsukuba, 1-2 Kasuga, Tsukuba, Ibaraki 305-8550 Japan}
\author{A. S. Zibrov}
\affiliation{Department of Physics,	Harvard University,	17 Oxford Street, Cambridge, MA 02138, USA}
\author{F. Jelezko}
\affiliation{Institute for Quantum Optics, University Ulm, Albert-Einstein-Allee 11, 89081 Ulm, Germany}
\author{M. D. Lukin}
\affiliation{Department of Physics,	Harvard University,	17 Oxford Street, Cambridge, MA 02138, USA}

\begin{abstract}
We demonstrate that silicon-vacancy  ($\siv$) centers in diamond can be used to efficiently generate coherent optical photons with excellent spectral properties.  
We show that these features are due to the inversion symmetry associated with $\siv$ centers, and demonstrate generation of indistinguishable single photons from separate emitters in a Hong-Ou-Mandel (HOM) interference experiment.  
Prospects for realizing efficient quantum network nodes using $\siv$ centers are discussed. 
\end{abstract}


\maketitle

The realization of quantum networks, in which local quantum processing nodes are connected over long distances via optical photons, is an outstanding challenge in quantum information science\cite{kimble2008quantum}.  
Over the past few years, atom-like systems in the solid state have emerged as a promising platform for achieving this goal.  
Key building blocks have been demonstrated using nitrogen-vacancy ($\nv$) centers in diamond, including long lived qubit memory\cite{maurer2012room}, spin-photon\cite{togan2010quantum} and spin-spin entanglement\cite{bernien2013heralded}, as well as teleportation between distant stationary qubits\cite{pfaff2014unconditional}.  
While $\nv$ centers can be used as excellent quantum registers, the current efforts to scale up these proof-of-concept experiments are limited by the small probability of coherent photon emission from $\nv$ centers and their spectral stability\cite{faraon2012coupling,chu2014coherent}.  
 Here we demonstrate that silicon-vacancy  ($\siv$) centers in diamond can be used to efficiently generate coherent optical photons with excellent spectral stability.  
We show that these features are due to the inversion symmetry associated with $\siv$ centers, and demonstrate generation of indistinguishable single photons from separate emitters in a Hong-Ou-Mandel (HOM) interference experiment \cite{hong1987measurement}.

\begin{figure}
\includegraphics[width=\linewidth]{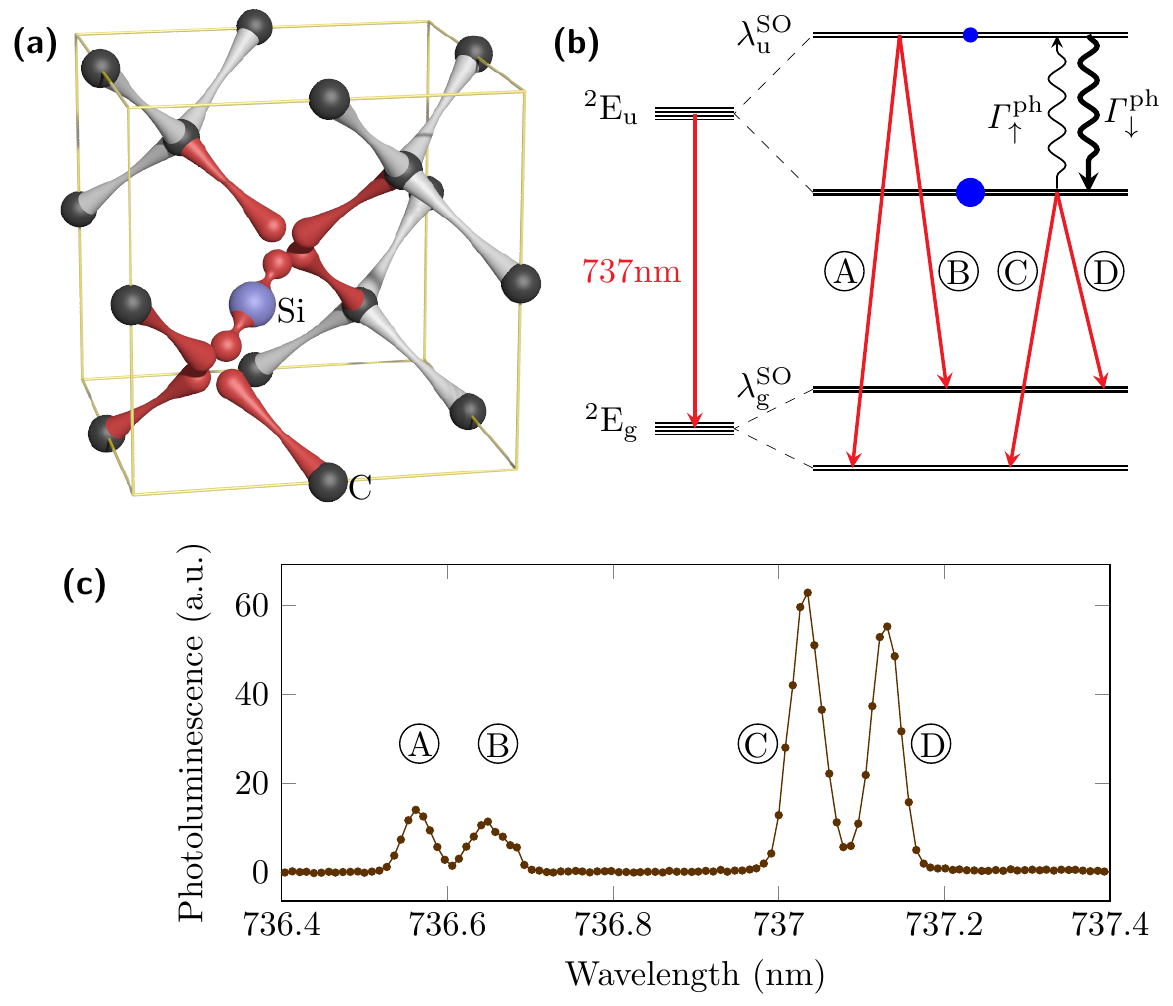}
\caption{
Electronic structure and optical transitions of the $\siv$ center.  
(a) The center is aligned along a $\langle 111 \rangle$ axis of the diamond host crystal, with the Silicon atom (Si) located in the middle of two empty lattice sites.
The system has $\Dthreed$ symmetry which includes inversion symmetry.  
(b) The optical transition is between different parity states, $\doubEu$ and $\doubEg$.  
Spin orbit interaction ($\lambdau\sim250$\,GHz, $\lambdag\sim50$\,GHz) partially lifts the degeneracy giving rise to doublets in the ground and excited states.  
Transitions A, B, $\transC$, D are all dipole allowed.  
(c) The emission spectrum measured using off-resonant excitation at 532\,nm on a single $\siv$ center at 4.5\,K.  
}
\label{structure}
\end{figure}

The negatively charged $\siv$ center in diamond consists of a silicon atom and a split vacancy as shown in \autoref{structure}(a) \cite{goss1996twelve,rogers2014electronic}.  
The silicon atom is centered between two empty lattice sites, and this 
$\Dthreed$
geometry forms an inversion symmetric potential for the electronic orbitals\cite{goss1996twelve}.
Recent measurements\cite{rogers2014electronic,hepp2014electronic} and first principle calculations\cite{gali2013ab} have contributed to a detailed understanding of the electronic structure of the $\siv$ center. 
As shown in \autoref{structure}(b), the ground and excited states each consist of a fourfold degenerate manifold where two degenerate orbitals are occupied by a $S=1/2$ particle\cite{muller2014optical}.
At zero magnetic field, the degeneracy is partially lifted by the spin-orbit interaction.  
Each excited state has dipole transitions to the two ground states forming an optical $\Lambda$ system, resulting in the emission spectrum shown in \autoref{structure}(c). 
These four transitions comprise the zero-phonon line (ZPL), which contains more than $70\%$ of the total fluorescence.
Remarkably, as discussed below, the inversion symmetry results in weak coupling of the ZPL transitions to charge fluctuations in the $\siv$ environment.
This leads to the absence of spectral diffusion\cite{rogers2013multiple} and a narrow inhomogeneous distribution\cite{sternschulte1994luminescence}.

\begin{figure}
\includegraphics[width=\linewidth]{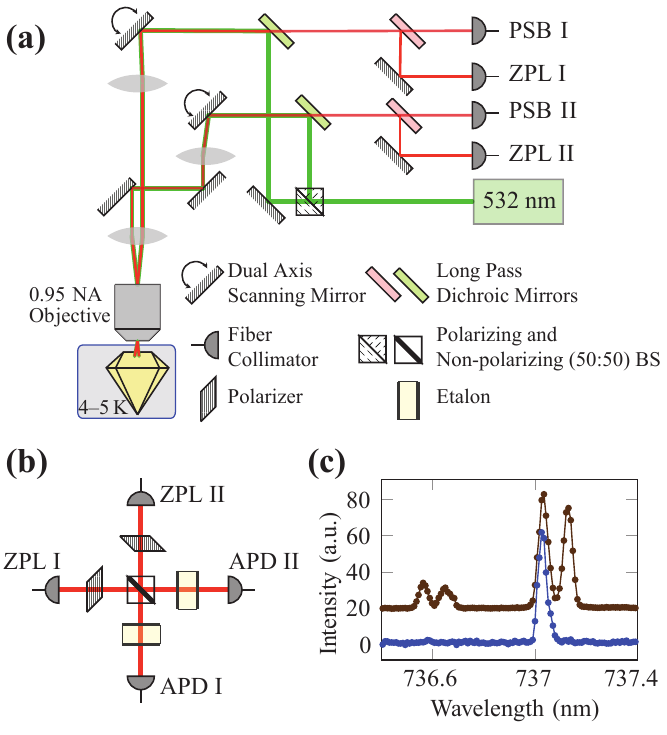}
\caption{
	Schematic of the two-channel confocal microscope built for the HOM experiment. 
	(a) Channels I and II were used to address different emitters separated by tens of micrometers in the same sample.
	A continuous-wave 532\,nm laser was used for excitation, and fluorescence was collected in single mode fibers on ZPL and PSB ports simultaneously.  
	(b) Collected ZPL fluorescence from the two channels were directed onto a free-space 50:50 non-polarizing beam splitter.  
	Linear polarizers were used to control the polarization of the single photons varying their distinguishability.  
	Etalons were used to filter transition $\transC$ before detection.  
	(c) Emission spectrum before (brown) and after the etalons (blue).  
}
\label{fig:setup}
\end{figure}

To demonstrate coherent emission of indistinguishable single photons from separate $\siv$ centers we use a Hong-Ou-Mandel interference experiment.  
The interference of two identical single photons impinging on a beamsplitter results in perfect photon bunching, with a vanishing probability of detecting coincident photons at the two different output ports.  
In our experiments two separate $\siv$ centers, cooled to cryogenic temperatures, were excited using a two-channel confocal optical microscope shown in \autoref{fig:setup}(a).  
Dichroic mirrors were used to simultaneously collect the $\siv$ fluorescence on both the ZPL ($\lambda \sim$ 737\,nm) and phonon-side-band (PSB, $\lambda \sim$ 760--860\,nm).
In order to isolate a single two-level transition, the emission spectrum was filtered by solid etalons (Figure 2(b)) with a free spectral range of 20\,GHz and a bandwidth of 1\,GHz.
The etalons were tuned by temperature to transition $\transC$ and the transmitted fluorescence spectrum is shown in  \autoref{fig:setup}(c), where only a single peak is visible as desired for indistinguishable photon generation.  

\begin{figure}
\includegraphics[width=\linewidth]{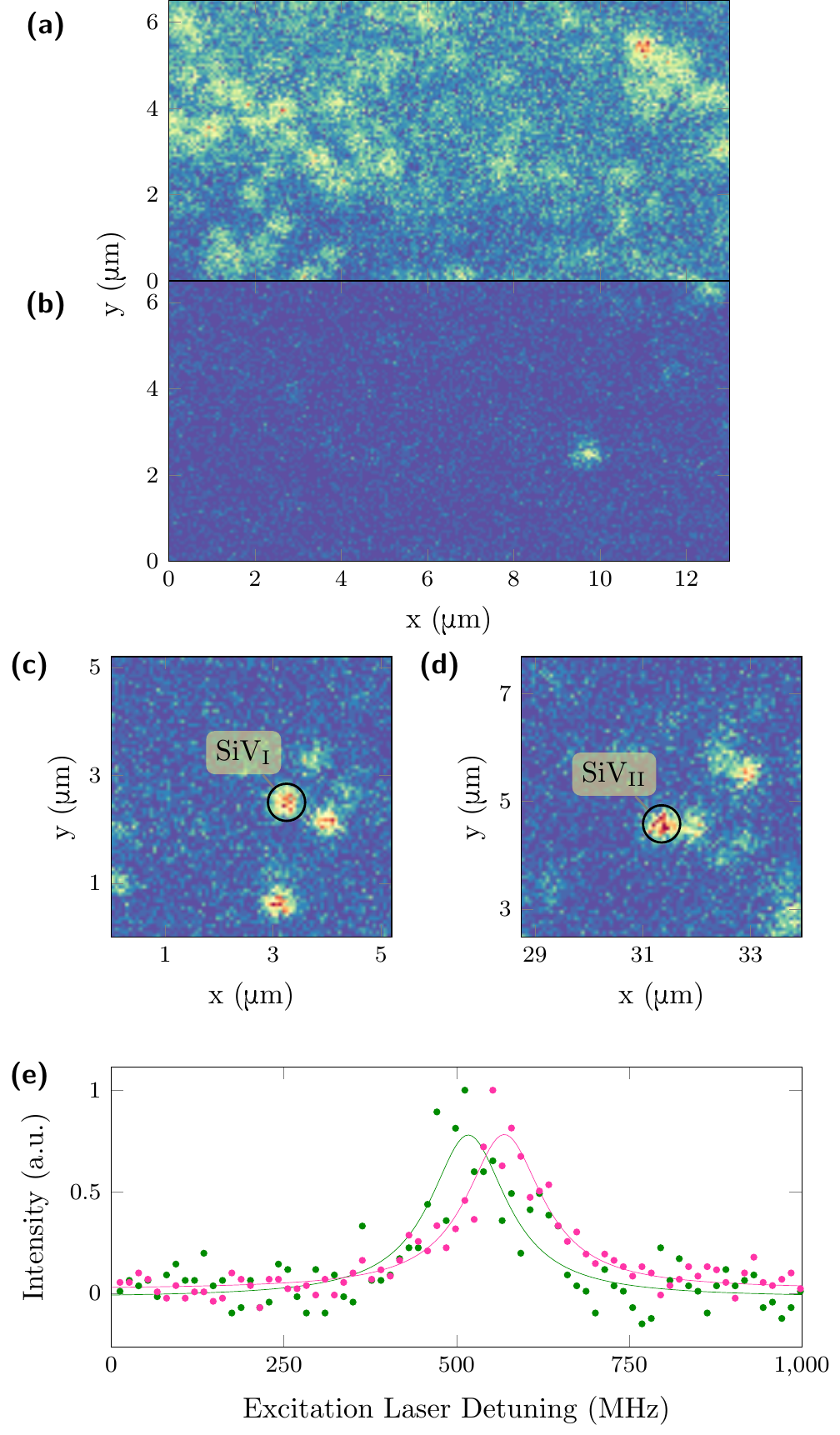}
\caption{
	Inhomogeneous distribution of $\siv$ centers.
	(a) The probe laser frequency was fixed to the ensemble average of $\nu_0=406.7001$\,THz for transition $\transC$ while scanning the sample.  
	A high density of resonant emitters is visible with a large background.
	(b) Scan of the same region with the laser tuned to $\nu_1=\nu_0+1.5$\,GHz.  
	Due to the narrow inhomogeneous distribution, only few resonant sites are visible and the background level is low. 
	(c) and (d) show the two emitters, $\siva$ and $\sivb$, used for the HOM interference experiment at frequency $\nu\sim\nu_1$. 
	(e) PLE spectrum for $\siva$ (green) and $\sivb$ (pink) with measured full width half maximum (FWHM) of 135.8 and 134.6 MHz respectively, and lines separated by 52.1\,MHz.  
}
\label{fig:density_comparison}
\end{figure}

To probe the inhomogeneous distribution (see Supp. Mat.\cite{SI}) within the sample and select spectrally overlapping sites, the emitters were resonantly excited with a 737\,nm probe laser using the ZPL.
The laser was tuned to the center frequency ($\nu_0$) of the inhomogeneous distribution for transition $\transC$ while monitoring fluorescence intensity in the PSB.  
\autoref{fig:density_comparison}(a) shows the diamond sample imaged by this technique in a region where the resonant site density was high, leading to a high background in any photon correlation experiments.  
In order to isolate single $\siv$ centers and minimize background from other emitters\cite{moerner1989optical}, the laser was tuned to the edge of the inhomogeneous distribution ($\nu_1$) in \autoref{fig:density_comparison}(b).  
Figures \ref{fig:density_comparison}(c,d) show the two emitters that were chosen for the HOM experiment at frequency $\nu \sim \nu_1$.  
The images in Figures \ref{fig:density_comparison}(c,d) were taken under 532\,nm excitation while detecting ZPL photons through the etalons.  
Here the frequency selectivity is limited by the bandwidth of the etalons ($\sim 1$\,GHz) and therefore more emitters are visible than in the resonant excitation scan (\autoref{fig:density_comparison}(b)).  
Photoluminescence excitation (PLE) spectra of the emitters, $\siva$ (green) and $\sivb$ (pink),  reveal transitions separated by 52.1\,MHz with full width half maximum (FWHM) of 135.8 and 134.6\,MHz respectively.  
For comparison, the lifetime of the excited states was measured to be $1.73\pm0.05$\,ns at temperatures below 50\,K corresponding to a transform limited linewidth of 94\,MHz.

\begin{figure}
\includegraphics[width=\linewidth]{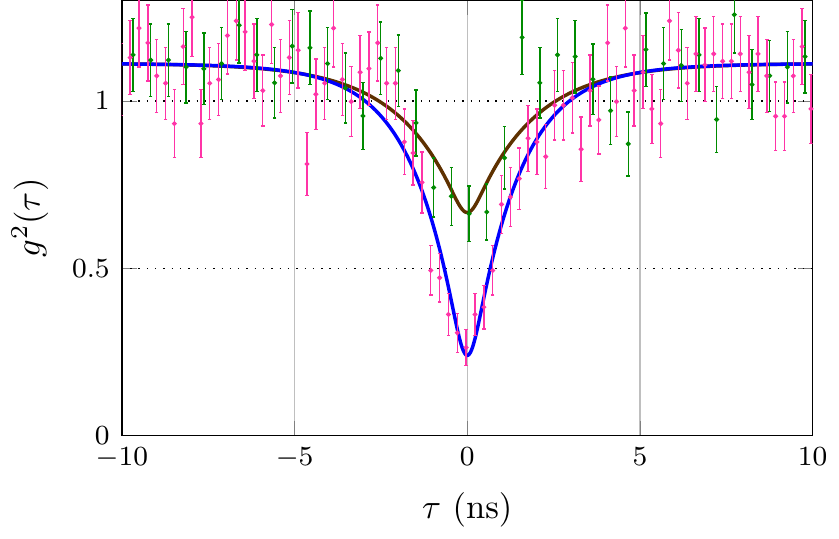}
\caption{
	Hong-Ou-Mandel interference experiment.  
	The second order intensity correlation function $g^2( \tau )$ is plotted for two cases: 
	(i) Pink data shows the results for indistinguishable single photons with identical polarizations, $g^2_\parallel (0)\!=\!0.26 \pm 0.05$.  
	The error bars denote shot noise estimates.  
	(ii) Green data shows the results when photons from one emitter are orthogonally polarized and hence distinguishable, $g^2_\perp (0) = 0.66 \pm 0.08$.
	The blue and brown solid lines represent our model using independently measured parameters, only fitting a single parameter for background events in both datasets.  
}
\label{fig:hom}
\end{figure}

For the HOM measurement, single photons emitted from $\siva$ and $\sivb$ on transition $\transC$ were directed to the input ports 1 and 2 of the beamsplitter respectively (see \autoref{fig:setup}(a,b)).  
\autoref{fig:hom} shows two measurements where the degree of indistinguishability of single photons is varied by changing the photon polarization. 
The two datasets show the second order intensity correlation function, $g^{2}(\tau)$, measured for indistinguishable (pink) and distinguishable (green) photon states.  
For identically polarized indistinguishable photons, we find $g^2_\parallel (0) = 0.26 \pm 0.05$ where the error bars denote shot noise estimates.  
After rotating the fluorescence polarization of $\sivb$ by $90^\circ$ to make the photon sources distinguishable, $g^2_\perp (0) = 0.66 \pm 0.08$ was observed.  
These results clearly demonstrate two-photon interference corresponding to a measured HOM visibility of $$\eta=\frac{g^2_\perp (0)}{g^2_\parallel (0) + g^2_\perp (0)} = 0.72 \pm 0.05.$$  
The time dynamics of $g^2(\tau)$ is understood via independent measurements of the excited state lifetime, absorption linewidth, and detector timing response.  
Our model (solid curves, see Supp. Mat.\cite{SI}) is in excellent agreement with the measured time dynamics, showing that the emitters were spectrally stable throughout the 4-hour acquisition period.
We find that the interference visibility, $\eta$, is limited by about equal contributions from detector timing response and background events.

We next turn to a discussion of the key properties of $\siv$ centers which made the present observations possible.
Despite uncertainty about the absolute quantum yield\cite{rogers2013multiple}, the strong ZPL of $\siv$ \cite{neu2011single} means that photons are emitted at high rates into the optical transition of interest.
Inhomogeneous broadening corresponded to only a few transition linewidths (see Supp. Mat.\cite{SI}), and high spectral stability of the transitions has been observed in bulk diamond\cite{rogers2013multiple} and nanodiamonds\cite{muller2014optical}. 
Together with these observations, our work shows that the optical coherence properties of $\siv$ centers can be superior to those of $\nv$ centers\cite{sipahigil2012quantum,chu2014coherent}. 
Some of this advantage can be understood to result from the inversion symmetry of $\siv$ centers (which reduces sensitivity to electric field). In addition, it is important to consider the effects of phonons (strain) resulting in homogenous (inhomogeneous) broadening mechanisms.

The electronic orbitals of the $\siv$ center are parity eigenstates due to the inversion symmetry of the defect.
The optical transitions take place between states of different parity, $\doubEg$ and $\doubEu$, which differ in phase but have similar charge densities\cite{gali2013ab}.
This small change in the electronic charge density results in the strong ZPL, since optical excitations do not couple efficiently to local vibrations.
The coherence of the optical transitions can also suffer from spectral diffusion, a time dependent change in the optical transition frequencies that results in an increased linewidth.
This effect is commonly observed for $\nv$ centers, where the dominant source of spectral diffusion has been shown to be from local electronic charge fluctuations\cite{siyushev2013optically}.
These changes in the charge environment result in a fluctuating electric field at the emitter which reduces the coherence of the optical transitions via DC Stark shift\cite{tamarat2006stark,chu2014coherent}. 
The sensitivity of the optical transition frequencies to electric field fluctuations depends on the permanent electric dipole moments of the orbital states of the emitter.
Since the electronic states of the $\siv$ center have vanishing permanent electric dipole moments due to their inversion symmetry, the optical transitions are relatively insensitive to external electric fields.
This protects the optical coherence from charge dynamics in the crystal, preventing spectral diffusion and narrowing the inhomogeneous distribution of transition frequencies.

Additional homogeneous and inhomogeneous broadening mechanisms are provided by phonons and strain.
Displacements of atoms in the host crystal can affect the optical transitions in two different ways.  
Static distortions, or strain, may reduce the symmetry of the defect and change the energy splittings\cite{sternschulte1994luminescence} shown in \autoref{structure}(b).
A variation in local strain contributes to the inhomogeneous distribution of the resonance frequencies\cite{SI}.  
Displacements of the atoms can also give rise to dynamic effects during an optical excitation cycle.  
Acoustic phonons have been shown to cause orbital relaxation between $\Ex$ and $\Ey$ states for the $\nv$ center in diamond\cite{fu2009observation}.  
For $\siv$ centers, a similar process can happen between excited state orbitals by absorption ($\Gammaup$) or emission ($\Gammadown$) of an acoustic phonon as shown in \autoref{structure}(b).  
Populations in the upper and lower excited state branches follow a Boltzmann distribution confirming thermalization of orbital states by phonons\cite{sternschulte1994luminescence, rogers2013multiple}.  
At low temperatures ($k_\mathrm{B} T \ll \hbar \lambdau \sim 250$\,GHz) spontaneous emission dominates over stimulated processes ($\Gammaup \ll \Gammadown$).  
To obtain an optical transition isolated from the phonon bath, our experiments were performed at 4.5--5\,K ($\sim 100$\,GHz) using the lower excited state branch.  
At these temperatures, we estimate a thermal broadening on transition $\transC$ of about 12\,MHz \cite{rogers2013multiple}.

Our observations establish the $\siv$ center as an excellent source of indistinguishable single photons. 
A strong ZPL transition, narrow inhomogeneous distribution, and spectral stability combine to make it a promising platform for applications in the fields of quantum networks and long distance quantum communication.
In particular, it should be possible to integrate $\siv$ centers inside nanophotonic cavities\cite{burek2012free,hausmann2013coupling,riedrich2012one,lee2012coupling,faraon2012coupling} while maintaining their spectral properties owing to their insensitivity to electric fields.
This may allow the realization of GHz bandwidth deterministic single photon sources\cite{kuhn2002deterministic} and a broadband system for quantum nonlinear optics at the single photon level\cite{tiecke2014nanophotonic}.
The small inhomogeneous distribution also makes $\siv$ centers promising candidates as sources of multiple indistinguishable photons for linear optics quantum computing\cite{kok2007linear}. 
Furthermore, the spin degree of freedom in the ground state \cite{muller2014optical} can potentially be utilized to store quantum information, allowing the use of $\siv$ centers as quantum registers for quantum network applications \cite{childress2005fault}. 
Coupling to the ${}^{29}$Si nuclear spin via hyperfine interactions\cite{edmonds2008electron} might allow realization of long lived quantum memories\cite{maurer2012room}.
Beyond these specific applications, the symmetry arguments presented above suggest that inversion symmetry might play an important role in the identification of new centers with suitable properties for quantum information science and technology\cite{weber2010quantum}.

We thank J. D. Thompson, N. de Leon, A. Gali, M. L. Goldman, T. Zibrova for theoretical discussions and experimental help. This work was supported by NSF, CUA, and JSPS KAKENHI (No.26246001). A. S. and K. J. contributed equally to this work.

\section{Supplemental Material}

\subsection{Sample Information}
The $\siv$ sites were found by looking through a $\{001\}$ face on a Type IIa diamond.  
The diamond had a CVD layer grown on top of a low strain HPHT diamond substrate.  
The $\siv$ sites were incorporated in the growth process by etching a piece of silicon-carbide (SiC) with the growth plasma.  
This technique produced highly uniform centers with a small inhomogeneous distribution \cite{rogers2013multiple}. 

\subsection{Inhomogeneous distribution}

One of the key requirements for a Hong-Ou-Mandel (HOM) experiment using separate emitters in solids is to isolate emitters with identical emission frequencies. 
For emitters in solids, the inhomogeneous distribution is typically orders of magnitude larger than the optical transition linewidths, and large inhomogeneous distributions require some combination of postselection of emitters or active tuning\cite{lettow2010quantum,patel2010two,flagg2010interference,sipahigil2012quantum}. 
Owing to the insensitivity of its transitions to electric fields, $\siv$ has a much narrower inhomogeneous distribution as discussed in the main text.
\autoref{inhomo_dist} shows the measured inhomogeneous distribution using the technique described in Figure 3 of the manuscript.
The total fluorescence counts in a field of view of   $\sim150\,\mu$m${}^2$ are plotted for each laser excitation frequency $\nu$.
The measured inhomogeneous distribution width of $364.5 \pm 33.0$\,MHz is only a few times the lifetime limited linewidth of 94MHz, suggesting that most of the emitters in the field of view could be used for a HOM demonstration.

\makeatletter 
\renewcommand{\thefigure}{S\@arabic\c@figure}
\makeatother
\begin{figure}
\includegraphics[width=0.9\linewidth]{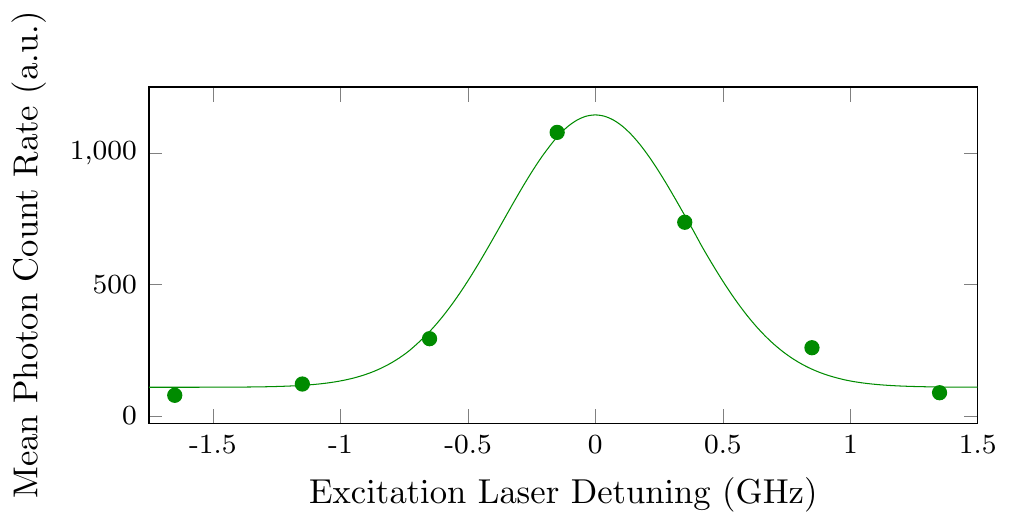}
\caption{
Inhomogeneous distribution of the $\siv$ resonance.  
A Gaussian fit to the distribution reveals a distribution with a width of $\sigma=364.5 \pm 33.0$\,MHz.  
The shot noise error was estimated to be smaller than the data points.
}
\label{inhomo_dist}
\end{figure}

Since the $\siv$ centers couple to strain and phonons but not to electric fields, they could be used as a nanoscale sensor of the strain distribution in a crystal.
For example, the bright emitter in Figure 3(b) was on the wings of the inhomogeneous distribution, possibly due to a point defect near the emitter that shifts its frequency substantially with respect to the rest of the ensemble. 
%
%

\subsection{Time dynamics of $g^2( \tau )$}

\begin{figure}
\includegraphics[width=0.9\linewidth]{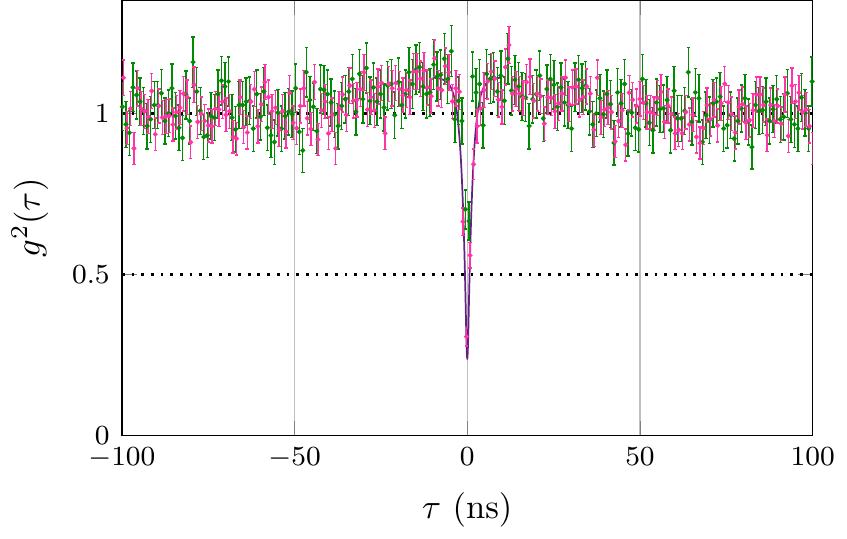}
\caption{
Normalization for the HOM results.  
The long time dynamics of the second order intensity correlation function $g^2( \tau )$ are shown where the data was coarse binned with 1024 ps intervals.
The $g^2( \tau )$ function was normalized using the steady state coincidence levels at long time delays where effects of bunching\cite{neu2012photophysics} for $|\tau| >\,50$ ns vanish.
}
\label{fig:homLong}
\end{figure}

Here we describe the solid curves in Figure 4 of the main text that represents our model. 
The shape of the measured second order correlation function, $g^2( \tau )$, for short time delays can be explained using the expression
 $$g^2( \tau )=
			 \,\frac{1}{2} g^{(2)}_{1}(\tau) 
			+ \frac{1}{2} \left(1	- \chi \,|g^{(1)}_{1}(\tau)|^2	 \cos(\Delta \omega)	\,\right)\,,$$
where $g^{(1)}_1$ and $g^{(2)}_1$ are the field (first order) and intensity (second order) autocorrelation functions for a single emitter. $\Delta$ is the detuning between the two emitters, $\chi= 1 (0)$ for indistiguishable(distinguishable) photons
\cite{lettow2010quantum,sipahigil2012quantum}.
We substitute $g^{(2)}_1(\tau)= 1- (1 - c_\mathrm{B})\, \exp\left(-\frac{|\tau|}{\tau_0}\right)$ and $|g^{(1)}_1(\tau)|^2= (1 - c_\mathrm{B})\, \exp \left(-\frac{|\tau|}{\tau_c}\right)$ where $\tau_0=1.73$ ns is the excited state lifetime and $\tau_c=1.18$ ns is the coherence time of the transition calculated from the linewidths of the PLE spectra.
The only free parameter in this model is the ratio of the background events to total events, $c_\mathrm{B}$,  which we consistently find to be 0.12 from the fits to control and HOM experiments.
In our model we also take into account the effects of detector timing jitter which was characterized using a pulsed laser.
The timing jitter of our detectors can be modeled as a gaussian with $\sigma\sim150$\,ps for each detector (PicoQuant $\tau$-SPAD).
To fit the HOM data, we used an analytical expression which is a convolution of the expression above with the detector timing response.
The model and the data were normalized such that $g^2( |\tau| \rightarrow \infty)=1$ as shown in \autoref{fig:homLong} to take the weak bunching effect into account.
The bunching effect is related to the population dynamics of a single emitter and hints at the presence of a metastable state \cite{neu2012photophysics}.

\bibliography{hom2}

\begin{thebibliography}{36}%
\makeatletter
\providecommand \@ifxundefined [1]{%
 \@ifx{#1\undefined}
}%
\providecommand \@ifnum [1]{%
 \ifnum #1\expandafter \@firstoftwo
 \else \expandafter \@secondoftwo
 \fi
}%
\providecommand \@ifx [1]{%
 \ifx #1\expandafter \@firstoftwo
 \else \expandafter \@secondoftwo
 \fi
}%
\providecommand \natexlab [1]{#1}%
\providecommand \enquote  [1]{``#1''}%
\providecommand \bibnamefont  [1]{#1}%
\providecommand \bibfnamefont [1]{#1}%
\providecommand \citenamefont [1]{#1}%
\providecommand \href@noop [0]{\@secondoftwo}%
\providecommand \href [0]{\begingroup \@sanitize@url \@href}%
\providecommand \@href[1]{\@@startlink{#1}\@@href}%
\providecommand \@@href[1]{\endgroup#1\@@endlink}%
\providecommand \@sanitize@url [0]{\catcode `\\12\catcode `\$12\catcode
  `\&12\catcode `\#12\catcode `\^12\catcode `\_12\catcode `\%12\relax}%
\providecommand \@@startlink[1]{}%
\providecommand \@@endlink[0]{}%
\providecommand \url  [0]{\begingroup\@sanitize@url \@url }%
\providecommand \@url [1]{\endgroup\@href {#1}{\urlprefix }}%
\providecommand \urlprefix  [0]{URL }%
\providecommand \Eprint [0]{\href }%
\providecommand \doibase [0]{http://dx.doi.org/}%
\providecommand \selectlanguage [0]{\@gobble}%
\providecommand \bibinfo  [0]{\@secondoftwo}%
\providecommand \bibfield  [0]{\@secondoftwo}%
\providecommand \translation [1]{[#1]}%
\providecommand \BibitemOpen [0]{}%
\providecommand \bibitemStop [0]{}%
\providecommand \bibitemNoStop [0]{.\EOS\space}%
\providecommand \EOS [0]{\spacefactor3000\relax}%
\providecommand \BibitemShut  [1]{\csname bibitem#1\endcsname}%
\let\auto@bib@innerbib\@empty
\bibitem [{\citenamefont {Kimble}(2008)}]{kimble2008quantum}%
  \BibitemOpen
  \bibfield  {author} {\bibinfo {author} {\bibfnamefont {H.~J.}\ \bibnamefont
  {Kimble}},\ }\href
  {http://www.nature.com/nature/journal/v453/n7198/abs/nature07127.html}
  {\bibfield  {journal} {\bibinfo  {journal} {Nature (London)}\ }\textbf
  {\bibinfo {volume} {453}},\ \bibinfo {pages} {1023} (\bibinfo {year}
  {2008})}\BibitemShut {NoStop}%
\bibitem [{\citenamefont {Maurer}\ \emph {et~al.}(2012)\citenamefont {Maurer},
  \citenamefont {Kucsko}, \citenamefont {Latta}, \citenamefont {Jiang},
  \citenamefont {Yao}, \citenamefont {Bennett}, \citenamefont {Pastawski},
  \citenamefont {Hunger}, \citenamefont {Chisholm}, \citenamefont {Markham}
  \emph {et~al.}}]{maurer2012room}%
  \BibitemOpen
  \bibfield  {author} {\bibinfo {author} {\bibfnamefont {P.~C.}\ \bibnamefont
  {Maurer}}, \bibinfo {author} {\bibfnamefont {G.}~\bibnamefont {Kucsko}},
  \bibinfo {author} {\bibfnamefont {C.}~\bibnamefont {Latta}}, \bibinfo
  {author} {\bibfnamefont {L.}~\bibnamefont {Jiang}}, \bibinfo {author}
  {\bibfnamefont {N.~Y.}\ \bibnamefont {Yao}}, \bibinfo {author} {\bibfnamefont
  {S.~D.}\ \bibnamefont {Bennett}}, \bibinfo {author} {\bibfnamefont
  {F.}~\bibnamefont {Pastawski}}, \bibinfo {author} {\bibfnamefont
  {D.}~\bibnamefont {Hunger}}, \bibinfo {author} {\bibfnamefont
  {N.}~\bibnamefont {Chisholm}}, \bibinfo {author} {\bibfnamefont
  {M.}~\bibnamefont {Markham}},  \emph {et~al.},\ }\href
  {http://www.sciencemag.org/content/336/6086/1283} {\bibfield  {journal}
  {\bibinfo  {journal} {Science}\ }\textbf {\bibinfo {volume} {336}},\ \bibinfo
  {pages} {1283} (\bibinfo {year} {2012})}\BibitemShut {NoStop}%
\bibitem [{\citenamefont {Togan}\ \emph {et~al.}(2010)\citenamefont {Togan},
  \citenamefont {Chu}, \citenamefont {Trifonov}, \citenamefont {Jiang},
  \citenamefont {Maze}, \citenamefont {Childress}, \citenamefont {Dutt},
  \citenamefont {S{\o}rensen}, \citenamefont {Hemmer}, \citenamefont {Zibrov}
  \emph {et~al.}}]{togan2010quantum}%
  \BibitemOpen
  \bibfield  {author} {\bibinfo {author} {\bibfnamefont {E.}~\bibnamefont
  {Togan}}, \bibinfo {author} {\bibfnamefont {Y.}~\bibnamefont {Chu}}, \bibinfo
  {author} {\bibfnamefont {A.~S.}\ \bibnamefont {Trifonov}}, \bibinfo {author}
  {\bibfnamefont {L.}~\bibnamefont {Jiang}}, \bibinfo {author} {\bibfnamefont
  {J.}~\bibnamefont {Maze}}, \bibinfo {author} {\bibfnamefont {L.}~\bibnamefont
  {Childress}}, \bibinfo {author} {\bibfnamefont {M.~G.}\ \bibnamefont {Dutt}},
  \bibinfo {author} {\bibfnamefont {A.~S.}\ \bibnamefont {S{\o}rensen}},
  \bibinfo {author} {\bibfnamefont {P.}~\bibnamefont {Hemmer}}, \bibinfo
  {author} {\bibfnamefont {A.}~\bibnamefont {Zibrov}},  \emph {et~al.},\
  }\href@noop {} {\bibfield  {journal} {\bibinfo  {journal} {Nature (London)}\
  }\textbf {\bibinfo {volume} {466}},\ \bibinfo {pages} {730} (\bibinfo {year}
  {2010})}\BibitemShut {NoStop}%
\bibitem [{\citenamefont {Bernien}\ \emph {et~al.}(2013)\citenamefont
  {Bernien}, \citenamefont {Hensen}, \citenamefont {Pfaff}, \citenamefont
  {Koolstra}, \citenamefont {Blok}, \citenamefont {Robledo}, \citenamefont
  {Taminiau}, \citenamefont {Markham}, \citenamefont {Twitchen}, \citenamefont
  {Childress} \emph {et~al.}}]{bernien2013heralded}%
  \BibitemOpen
  \bibfield  {author} {\bibinfo {author} {\bibfnamefont {H.}~\bibnamefont
  {Bernien}}, \bibinfo {author} {\bibfnamefont {B.}~\bibnamefont {Hensen}},
  \bibinfo {author} {\bibfnamefont {W.}~\bibnamefont {Pfaff}}, \bibinfo
  {author} {\bibfnamefont {G.}~\bibnamefont {Koolstra}}, \bibinfo {author}
  {\bibfnamefont {M.}~\bibnamefont {Blok}}, \bibinfo {author} {\bibfnamefont
  {L.}~\bibnamefont {Robledo}}, \bibinfo {author} {\bibfnamefont
  {T.}~\bibnamefont {Taminiau}}, \bibinfo {author} {\bibfnamefont
  {M.}~\bibnamefont {Markham}}, \bibinfo {author} {\bibfnamefont
  {D.}~\bibnamefont {Twitchen}}, \bibinfo {author} {\bibfnamefont
  {L.}~\bibnamefont {Childress}},  \emph {et~al.},\ }\href@noop {} {\bibfield
  {journal} {\bibinfo  {journal} {Nature (London)}\ }\textbf {\bibinfo {volume}
  {497}},\ \bibinfo {pages} {86} (\bibinfo {year} {2013})}\BibitemShut
  {NoStop}%
\bibitem [{\citenamefont {Pfaff}\ \emph {et~al.}(2014)\citenamefont {Pfaff},
  \citenamefont {Hensen}, \citenamefont {Bernien}, \citenamefont {van Dam},
  \citenamefont {Blok}, \citenamefont {Taminiau}, \citenamefont {Tiggelman},
  \citenamefont {Schouten}, \citenamefont {Markham}, \citenamefont {Twitchen}
  \emph {et~al.}}]{pfaff2014unconditional}%
  \BibitemOpen
  \bibfield  {author} {\bibinfo {author} {\bibfnamefont {W.}~\bibnamefont
  {Pfaff}}, \bibinfo {author} {\bibfnamefont {B.}~\bibnamefont {Hensen}},
  \bibinfo {author} {\bibfnamefont {H.}~\bibnamefont {Bernien}}, \bibinfo
  {author} {\bibfnamefont {S.~B.}\ \bibnamefont {van Dam}}, \bibinfo {author}
  {\bibfnamefont {M.}~\bibnamefont {Blok}}, \bibinfo {author} {\bibfnamefont
  {T.~H.}\ \bibnamefont {Taminiau}}, \bibinfo {author} {\bibfnamefont {M.~J.}\
  \bibnamefont {Tiggelman}}, \bibinfo {author} {\bibfnamefont {R.~N.}\
  \bibnamefont {Schouten}}, \bibinfo {author} {\bibfnamefont {M.}~\bibnamefont
  {Markham}}, \bibinfo {author} {\bibfnamefont {D.~J.}\ \bibnamefont
  {Twitchen}},  \emph {et~al.},\ }\href@noop {} {\bibfield  {journal} {\bibinfo
   {journal} {Science}\ ,\ \bibinfo {pages} {1253512}} (\bibinfo {year}
  {2014})}\BibitemShut {NoStop}%
\bibitem [{\citenamefont {Faraon}\ \emph {et~al.}(2012)\citenamefont {Faraon},
  \citenamefont {Santori}, \citenamefont {Huang}, \citenamefont {Acosta},\ and\
  \citenamefont {Beausoleil}}]{faraon2012coupling}%
  \BibitemOpen
  \bibfield  {author} {\bibinfo {author} {\bibfnamefont {A.}~\bibnamefont
  {Faraon}}, \bibinfo {author} {\bibfnamefont {C.}~\bibnamefont {Santori}},
  \bibinfo {author} {\bibfnamefont {Z.}~\bibnamefont {Huang}}, \bibinfo
  {author} {\bibfnamefont {V.~M.}\ \bibnamefont {Acosta}}, \ and\ \bibinfo
  {author} {\bibfnamefont {R.~G.}\ \bibnamefont {Beausoleil}},\ }\href@noop {}
  {\bibfield  {journal} {\bibinfo  {journal} {Phys. Rev. Lett.}\ }\textbf
  {\bibinfo {volume} {109}},\ \bibinfo {pages} {033604} (\bibinfo {year}
  {2012})}\BibitemShut {NoStop}%
\bibitem [{\citenamefont {Chu}\ \emph {et~al.}(2014)\citenamefont {Chu},
  \citenamefont {de~Leon}, \citenamefont {Shields}, \citenamefont {Hausmann},
  \citenamefont {Evans}, \citenamefont {Togan}, \citenamefont {Burek},
  \citenamefont {Markham}, \citenamefont {Stacey}, \citenamefont {Zibrov} \emph
  {et~al.}}]{chu2014coherent}%
  \BibitemOpen
  \bibfield  {author} {\bibinfo {author} {\bibfnamefont {Y.}~\bibnamefont
  {Chu}}, \bibinfo {author} {\bibfnamefont {N.}~\bibnamefont {de~Leon}},
  \bibinfo {author} {\bibfnamefont {B.}~\bibnamefont {Shields}}, \bibinfo
  {author} {\bibfnamefont {B.}~\bibnamefont {Hausmann}}, \bibinfo {author}
  {\bibfnamefont {R.}~\bibnamefont {Evans}}, \bibinfo {author} {\bibfnamefont
  {E.}~\bibnamefont {Togan}}, \bibinfo {author} {\bibfnamefont {M.~J.}\
  \bibnamefont {Burek}}, \bibinfo {author} {\bibfnamefont {M.}~\bibnamefont
  {Markham}}, \bibinfo {author} {\bibfnamefont {A.}~\bibnamefont {Stacey}},
  \bibinfo {author} {\bibfnamefont {A.}~\bibnamefont {Zibrov}},  \emph
  {et~al.},\ }\href@noop {} {\bibfield  {journal} {\bibinfo  {journal} {Nano
  Lett.}\ }\textbf {\bibinfo {volume} {14}},\ \bibinfo {pages} {1982} (\bibinfo
  {year} {2014})}\BibitemShut {NoStop}%
\bibitem [{\citenamefont {Hong}\ \emph {et~al.}(1987)\citenamefont {Hong},
  \citenamefont {Ou},\ and\ \citenamefont {Mandel}}]{hong1987measurement}%
  \BibitemOpen
  \bibfield  {author} {\bibinfo {author} {\bibfnamefont {C.~K.}\ \bibnamefont
  {Hong}}, \bibinfo {author} {\bibfnamefont {Z.~Y.}\ \bibnamefont {Ou}}, \ and\
  \bibinfo {author} {\bibfnamefont {L.}~\bibnamefont {Mandel}},\ }\href
  {\doibase 10.1103/PhysRevLett.59.2044} {\bibfield  {journal} {\bibinfo
  {journal} {Phys. Rev. Lett.}\ }\textbf {\bibinfo {volume} {59}},\ \bibinfo
  {pages} {2044} (\bibinfo {year} {1987})}\BibitemShut {NoStop}%
\bibitem [{\citenamefont {Goss}\ \emph {et~al.}(1996)\citenamefont {Goss},
  \citenamefont {Jones}, \citenamefont {Breuer}, \citenamefont {Briddon},\ and\
  \citenamefont {{\"O}berg}}]{goss1996twelve}%
  \BibitemOpen
  \bibfield  {author} {\bibinfo {author} {\bibfnamefont {J.~P.}\ \bibnamefont
  {Goss}}, \bibinfo {author} {\bibfnamefont {R.}~\bibnamefont {Jones}},
  \bibinfo {author} {\bibfnamefont {S.~J.}\ \bibnamefont {Breuer}}, \bibinfo
  {author} {\bibfnamefont {P.~R.}\ \bibnamefont {Briddon}}, \ and\ \bibinfo
  {author} {\bibfnamefont {S.}~\bibnamefont {{\"O}berg}},\ }\href
  {http://link.aps.org/doi/10.1103/PhysRevLett.77.3041} {\bibfield  {journal}
  {\bibinfo  {journal} {Phys. Rev. Lett.}\ }\textbf {\bibinfo {volume} {77}},\
  \bibinfo {pages} {3041} (\bibinfo {year} {1996})}\BibitemShut {NoStop}%
\bibitem [{\citenamefont {Rogers}\ \emph {et~al.}(2014)\citenamefont {Rogers},
  \citenamefont {Jahnke}, \citenamefont {Doherty}, \citenamefont {Dietrich},
  \citenamefont {McGuinness}, \citenamefont {M\"uller}, \citenamefont {Teraji},
  \citenamefont {Sumiya}, \citenamefont {Isoya}, \citenamefont {Manson},\ and\
  \citenamefont {Jelezko}}]{rogers2014electronic}%
  \BibitemOpen
  \bibfield  {author} {\bibinfo {author} {\bibfnamefont {L.~J.}\ \bibnamefont
  {Rogers}}, \bibinfo {author} {\bibfnamefont {K.~D.}\ \bibnamefont {Jahnke}},
  \bibinfo {author} {\bibfnamefont {M.~W.}\ \bibnamefont {Doherty}}, \bibinfo
  {author} {\bibfnamefont {A.}~\bibnamefont {Dietrich}}, \bibinfo {author}
  {\bibfnamefont {L.~P.}\ \bibnamefont {McGuinness}}, \bibinfo {author}
  {\bibfnamefont {C.}~\bibnamefont {M\"uller}}, \bibinfo {author}
  {\bibfnamefont {T.}~\bibnamefont {Teraji}}, \bibinfo {author} {\bibfnamefont
  {H.}~\bibnamefont {Sumiya}}, \bibinfo {author} {\bibfnamefont
  {J.}~\bibnamefont {Isoya}}, \bibinfo {author} {\bibfnamefont {N.~B.}\
  \bibnamefont {Manson}}, \ and\ \bibinfo {author} {\bibfnamefont
  {F.}~\bibnamefont {Jelezko}},\ }\href {\doibase 10.1103/PhysRevB.89.235101}
  {\bibfield  {journal} {\bibinfo  {journal} {Phys. Rev. B}\ }\textbf {\bibinfo
  {volume} {89}},\ \bibinfo {pages} {235101} (\bibinfo {year}
  {2014})}\BibitemShut {NoStop}%
\bibitem [{\citenamefont {Hepp}\ \emph {et~al.}(2014)\citenamefont {Hepp},
  \citenamefont {M{\"u}ller}, \citenamefont {Waselowski}, \citenamefont
  {Becker}, \citenamefont {Pingault}, \citenamefont {Sternschulte},
  \citenamefont {Steinm{\"u}ller-Nethl}, \citenamefont {Gali}, \citenamefont
  {Maze}, \citenamefont {Atat{\"u}re},\ and\ \citenamefont
  {Becher}}]{hepp2014electronic}%
  \BibitemOpen
  \bibfield  {author} {\bibinfo {author} {\bibfnamefont {C.}~\bibnamefont
  {Hepp}}, \bibinfo {author} {\bibfnamefont {T.}~\bibnamefont {M{\"u}ller}},
  \bibinfo {author} {\bibfnamefont {V.}~\bibnamefont {Waselowski}}, \bibinfo
  {author} {\bibfnamefont {J.~N.}\ \bibnamefont {Becker}}, \bibinfo {author}
  {\bibfnamefont {B.}~\bibnamefont {Pingault}}, \bibinfo {author}
  {\bibfnamefont {H.}~\bibnamefont {Sternschulte}}, \bibinfo {author}
  {\bibfnamefont {D.}~\bibnamefont {Steinm{\"u}ller-Nethl}}, \bibinfo {author}
  {\bibfnamefont {A.}~\bibnamefont {Gali}}, \bibinfo {author} {\bibfnamefont
  {J.~R.}\ \bibnamefont {Maze}}, \bibinfo {author} {\bibfnamefont
  {M.}~\bibnamefont {Atat{\"u}re}}, \ and\ \bibinfo {author} {\bibfnamefont
  {C.}~\bibnamefont {Becher}},\ }\href {\doibase
  10.1103/PhysRevLett.112.036405} {\bibfield  {journal} {\bibinfo  {journal}
  {Phys. Rev. Lett.}\ }\textbf {\bibinfo {volume} {112}},\ \bibinfo {pages}
  {036405} (\bibinfo {year} {2014})}\BibitemShut {NoStop}%
\bibitem [{\citenamefont {Gali}\ and\ \citenamefont {Maze}(2013)}]{gali2013ab}%
  \BibitemOpen
  \bibfield  {author} {\bibinfo {author} {\bibfnamefont {A.}~\bibnamefont
  {Gali}}\ and\ \bibinfo {author} {\bibfnamefont {J.~R.}\ \bibnamefont
  {Maze}},\ }\href {\doibase 10.1103/PhysRevB.88.235205} {\bibfield  {journal}
  {\bibinfo  {journal} {Phys. Rev. B}\ }\textbf {\bibinfo {volume} {88}},\
  \bibinfo {pages} {235205} (\bibinfo {year} {2013})}\BibitemShut {NoStop}%
\bibitem [{\citenamefont {M{\"u}ller}\ \emph {et~al.}(2014)\citenamefont
  {M{\"u}ller}, \citenamefont {Hepp}, \citenamefont {Pingault}, \citenamefont
  {Neu}, \citenamefont {Gsell}, \citenamefont {Schreck}, \citenamefont
  {Sternschulte}, \citenamefont {Steinm{\"u}ller-Nethl}, \citenamefont
  {Becher},\ and\ \citenamefont {Atat{\"u}re}}]{muller2014optical}%
  \BibitemOpen
  \bibfield  {author} {\bibinfo {author} {\bibfnamefont {T.}~\bibnamefont
  {M{\"u}ller}}, \bibinfo {author} {\bibfnamefont {C.}~\bibnamefont {Hepp}},
  \bibinfo {author} {\bibfnamefont {B.}~\bibnamefont {Pingault}}, \bibinfo
  {author} {\bibfnamefont {E.}~\bibnamefont {Neu}}, \bibinfo {author}
  {\bibfnamefont {S.}~\bibnamefont {Gsell}}, \bibinfo {author} {\bibfnamefont
  {M.}~\bibnamefont {Schreck}}, \bibinfo {author} {\bibfnamefont
  {H.}~\bibnamefont {Sternschulte}}, \bibinfo {author} {\bibfnamefont
  {D.}~\bibnamefont {Steinm{\"u}ller-Nethl}}, \bibinfo {author} {\bibfnamefont
  {C.}~\bibnamefont {Becher}}, \ and\ \bibinfo {author} {\bibfnamefont
  {M.}~\bibnamefont {Atat{\"u}re}},\ }\href@noop {} {\bibfield  {journal}
  {\bibinfo  {journal} {Nat. Commun.}\ }\textbf {\bibinfo {volume} {5}}
  (\bibinfo {year} {2014})}\BibitemShut {NoStop}%
\bibitem [{\citenamefont {Rogers}\ \emph {et~al.}(2013)\citenamefont {Rogers},
  \citenamefont {Jahnke}, \citenamefont {Marseglia}, \citenamefont
  {M{\"u}ller}, \citenamefont {Naydenov}, \citenamefont {Schauffert},
  \citenamefont {Kranz}, \citenamefont {Teraji}, \citenamefont {Isoya},
  \citenamefont {McGuinness} \emph {et~al.}}]{rogers2013multiple}%
  \BibitemOpen
  \bibfield  {author} {\bibinfo {author} {\bibfnamefont {L.~J.}\ \bibnamefont
  {Rogers}}, \bibinfo {author} {\bibfnamefont {K.~D.}\ \bibnamefont {Jahnke}},
  \bibinfo {author} {\bibfnamefont {L.}~\bibnamefont {Marseglia}}, \bibinfo
  {author} {\bibfnamefont {C.}~\bibnamefont {M{\"u}ller}}, \bibinfo {author}
  {\bibfnamefont {B.}~\bibnamefont {Naydenov}}, \bibinfo {author}
  {\bibfnamefont {H.}~\bibnamefont {Schauffert}}, \bibinfo {author}
  {\bibfnamefont {C.}~\bibnamefont {Kranz}}, \bibinfo {author} {\bibfnamefont
  {T.}~\bibnamefont {Teraji}}, \bibinfo {author} {\bibfnamefont
  {J.}~\bibnamefont {Isoya}}, \bibinfo {author} {\bibfnamefont {L.~P.}\
  \bibnamefont {McGuinness}},  \emph {et~al.},\ }\href@noop {} {\bibfield
  {journal} {\bibinfo  {journal} {arXiv preprint arXiv:1310.3804}\ } (\bibinfo
  {year} {2013})}\BibitemShut {NoStop}%
\bibitem [{\citenamefont {Sternschulte}\ \emph {et~al.}(1994)\citenamefont
  {Sternschulte}, \citenamefont {Thonke}, \citenamefont {Sauer}, \citenamefont
  {M{\"u}nzinger},\ and\ \citenamefont
  {Michler}}]{sternschulte1994luminescence}%
  \BibitemOpen
  \bibfield  {author} {\bibinfo {author} {\bibfnamefont {H.}~\bibnamefont
  {Sternschulte}}, \bibinfo {author} {\bibfnamefont {K.}~\bibnamefont
  {Thonke}}, \bibinfo {author} {\bibfnamefont {R.}~\bibnamefont {Sauer}},
  \bibinfo {author} {\bibfnamefont {P.~C.}\ \bibnamefont {M{\"u}nzinger}}, \
  and\ \bibinfo {author} {\bibfnamefont {P.}~\bibnamefont {Michler}},\
  }\href@noop {} {\bibfield  {journal} {\bibinfo  {journal} {Phys. Rev. B}\
  }\textbf {\bibinfo {volume} {50}},\ \bibinfo {pages} {14554} (\bibinfo {year}
  {1994})}\BibitemShut {NoStop}%
\bibitem [{SI()}]{SI}%
  \BibitemOpen
  \href@noop {} {}\bibinfo {note} {See Supplemental Material at [URL will be
  inserted by publisher] for more information on the inhomogeneous distribution
  and the model used for the HOM experiment.}\BibitemShut {Stop}%
\bibitem [{\citenamefont {Moerner}\ and\ \citenamefont
  {Kador}(1989)}]{moerner1989optical}%
  \BibitemOpen
  \bibfield  {author} {\bibinfo {author} {\bibfnamefont {W.~E.}\ \bibnamefont
  {Moerner}}\ and\ \bibinfo {author} {\bibfnamefont {L.}~\bibnamefont
  {Kador}},\ }\href@noop {} {\bibfield  {journal} {\bibinfo  {journal} {Phys.
  Rev. Lett.}\ }\textbf {\bibinfo {volume} {62}},\ \bibinfo {pages} {2535}
  (\bibinfo {year} {1989})}\BibitemShut {NoStop}%
\bibitem [{\citenamefont {Neu}\ \emph {et~al.}(2011)\citenamefont {Neu},
  \citenamefont {Steinmetz}, \citenamefont {Riedrich-M{\"o}ller}, \citenamefont
  {Gsell}, \citenamefont {Fischer}, \citenamefont {Schreck},\ and\
  \citenamefont {Becher}}]{neu2011single}%
  \BibitemOpen
  \bibfield  {author} {\bibinfo {author} {\bibfnamefont {E.}~\bibnamefont
  {Neu}}, \bibinfo {author} {\bibfnamefont {D.}~\bibnamefont {Steinmetz}},
  \bibinfo {author} {\bibfnamefont {J.}~\bibnamefont {Riedrich-M{\"o}ller}},
  \bibinfo {author} {\bibfnamefont {S.}~\bibnamefont {Gsell}}, \bibinfo
  {author} {\bibfnamefont {M.}~\bibnamefont {Fischer}}, \bibinfo {author}
  {\bibfnamefont {M.}~\bibnamefont {Schreck}}, \ and\ \bibinfo {author}
  {\bibfnamefont {C.}~\bibnamefont {Becher}},\ }\href@noop {} {\bibfield
  {journal} {\bibinfo  {journal} {New J. Phys.}\ }\textbf {\bibinfo {volume}
  {13}},\ \bibinfo {pages} {025012} (\bibinfo {year} {2011})}\BibitemShut
  {NoStop}%
\bibitem [{\citenamefont {Sipahigil}\ \emph {et~al.}(2012)\citenamefont
  {Sipahigil}, \citenamefont {Goldman}, \citenamefont {Togan}, \citenamefont
  {Chu}, \citenamefont {Markham}, \citenamefont {Twitchen}, \citenamefont
  {Zibrov}, \citenamefont {Kubanek},\ and\ \citenamefont
  {Lukin}}]{sipahigil2012quantum}%
  \BibitemOpen
  \bibfield  {author} {\bibinfo {author} {\bibfnamefont {A.}~\bibnamefont
  {Sipahigil}}, \bibinfo {author} {\bibfnamefont {M.~L.}\ \bibnamefont
  {Goldman}}, \bibinfo {author} {\bibfnamefont {E.}~\bibnamefont {Togan}},
  \bibinfo {author} {\bibfnamefont {Y.}~\bibnamefont {Chu}}, \bibinfo {author}
  {\bibfnamefont {M.}~\bibnamefont {Markham}}, \bibinfo {author} {\bibfnamefont
  {D.~J.}\ \bibnamefont {Twitchen}}, \bibinfo {author} {\bibfnamefont {A.~S.}\
  \bibnamefont {Zibrov}}, \bibinfo {author} {\bibfnamefont {A.}~\bibnamefont
  {Kubanek}}, \ and\ \bibinfo {author} {\bibfnamefont {M.~D.}\ \bibnamefont
  {Lukin}},\ }\href@noop {} {\bibfield  {journal} {\bibinfo  {journal} {Phys.
  Rev. Lett.}\ }\textbf {\bibinfo {volume} {108}},\ \bibinfo {pages} {143601}
  (\bibinfo {year} {2012})}\BibitemShut {NoStop}%
\bibitem [{\citenamefont {Siyushev}\ \emph {et~al.}(2013)\citenamefont
  {Siyushev}, \citenamefont {Pinto}, \citenamefont {V{\"o}r{\"o}s},
  \citenamefont {Gali}, \citenamefont {Jelezko},\ and\ \citenamefont
  {Wrachtrup}}]{siyushev2013optically}%
  \BibitemOpen
  \bibfield  {author} {\bibinfo {author} {\bibfnamefont {P.}~\bibnamefont
  {Siyushev}}, \bibinfo {author} {\bibfnamefont {H.}~\bibnamefont {Pinto}},
  \bibinfo {author} {\bibfnamefont {M.}~\bibnamefont {V{\"o}r{\"o}s}}, \bibinfo
  {author} {\bibfnamefont {A.}~\bibnamefont {Gali}}, \bibinfo {author}
  {\bibfnamefont {F.}~\bibnamefont {Jelezko}}, \ and\ \bibinfo {author}
  {\bibfnamefont {J.}~\bibnamefont {Wrachtrup}},\ }\href@noop {} {\bibfield
  {journal} {\bibinfo  {journal} {Phys. Rev. Lett.}\ }\textbf {\bibinfo
  {volume} {110}},\ \bibinfo {pages} {167402} (\bibinfo {year}
  {2013})}\BibitemShut {NoStop}%
\bibitem [{\citenamefont {Tamarat}\ \emph {et~al.}(2006)\citenamefont
  {Tamarat}, \citenamefont {Gaebel}, \citenamefont {Rabeau}, \citenamefont
  {Khan}, \citenamefont {Greentree}, \citenamefont {Wilson}, \citenamefont
  {Hollenberg}, \citenamefont {Prawer}, \citenamefont {Hemmer}, \citenamefont
  {Jelezko} \emph {et~al.}}]{tamarat2006stark}%
  \BibitemOpen
  \bibfield  {author} {\bibinfo {author} {\bibfnamefont {P.}~\bibnamefont
  {Tamarat}}, \bibinfo {author} {\bibfnamefont {T.}~\bibnamefont {Gaebel}},
  \bibinfo {author} {\bibfnamefont {J.}~\bibnamefont {Rabeau}}, \bibinfo
  {author} {\bibfnamefont {M.}~\bibnamefont {Khan}}, \bibinfo {author}
  {\bibfnamefont {A.}~\bibnamefont {Greentree}}, \bibinfo {author}
  {\bibfnamefont {H.}~\bibnamefont {Wilson}}, \bibinfo {author} {\bibfnamefont
  {L.}~\bibnamefont {Hollenberg}}, \bibinfo {author} {\bibfnamefont
  {S.}~\bibnamefont {Prawer}}, \bibinfo {author} {\bibfnamefont
  {P.}~\bibnamefont {Hemmer}}, \bibinfo {author} {\bibfnamefont
  {F.}~\bibnamefont {Jelezko}},  \emph {et~al.},\ }\href@noop {} {\bibfield
  {journal} {\bibinfo  {journal} {Phys. Rev. Lett.}\ }\textbf {\bibinfo
  {volume} {97}},\ \bibinfo {pages} {083002} (\bibinfo {year}
  {2006})}\BibitemShut {NoStop}%
\bibitem [{\citenamefont {Fu}\ \emph {et~al.}(2009)\citenamefont {Fu},
  \citenamefont {Santori}, \citenamefont {Barclay}, \citenamefont {Rogers},
  \citenamefont {Manson},\ and\ \citenamefont
  {Beausoleil}}]{fu2009observation}%
  \BibitemOpen
  \bibfield  {author} {\bibinfo {author} {\bibfnamefont {K.-M.~C.}\
  \bibnamefont {Fu}}, \bibinfo {author} {\bibfnamefont {C.}~\bibnamefont
  {Santori}}, \bibinfo {author} {\bibfnamefont {P.~E.}\ \bibnamefont
  {Barclay}}, \bibinfo {author} {\bibfnamefont {L.~J.}\ \bibnamefont {Rogers}},
  \bibinfo {author} {\bibfnamefont {N.~B.}\ \bibnamefont {Manson}}, \ and\
  \bibinfo {author} {\bibfnamefont {R.~G.}\ \bibnamefont {Beausoleil}},\
  }\href@noop {} {\bibfield  {journal} {\bibinfo  {journal} {Phys. Rev. Lett.}\
  }\textbf {\bibinfo {volume} {103}},\ \bibinfo {pages} {256404} (\bibinfo
  {year} {2009})}\BibitemShut {NoStop}%
\bibitem [{\citenamefont {Burek}\ \emph {et~al.}(2012)\citenamefont {Burek},
  \citenamefont {de~Leon}, \citenamefont {Shields}, \citenamefont {Hausmann},
  \citenamefont {Chu}, \citenamefont {Quan}, \citenamefont {Zibrov},
  \citenamefont {Park}, \citenamefont {Lukin},\ and\ \citenamefont {Lon{\v
  c}ar}}]{burek2012free}%
  \BibitemOpen
  \bibfield  {author} {\bibinfo {author} {\bibfnamefont {M.~J.}\ \bibnamefont
  {Burek}}, \bibinfo {author} {\bibfnamefont {N.~P.}\ \bibnamefont {de~Leon}},
  \bibinfo {author} {\bibfnamefont {B.~J.}\ \bibnamefont {Shields}}, \bibinfo
  {author} {\bibfnamefont {B.~J.}\ \bibnamefont {Hausmann}}, \bibinfo {author}
  {\bibfnamefont {Y.}~\bibnamefont {Chu}}, \bibinfo {author} {\bibfnamefont
  {Q.}~\bibnamefont {Quan}}, \bibinfo {author} {\bibfnamefont {A.~S.}\
  \bibnamefont {Zibrov}}, \bibinfo {author} {\bibfnamefont {H.}~\bibnamefont
  {Park}}, \bibinfo {author} {\bibfnamefont {M.~D.}\ \bibnamefont {Lukin}}, \
  and\ \bibinfo {author} {\bibfnamefont {M.}~\bibnamefont {Lon{\v c}ar}},\
  }\href@noop {} {\bibfield  {journal} {\bibinfo  {journal} {Nano Lett.}\
  }\textbf {\bibinfo {volume} {12}},\ \bibinfo {pages} {6084} (\bibinfo {year}
  {2012})}\BibitemShut {NoStop}%
\bibitem [{\citenamefont {Hausmann}\ \emph {et~al.}(2013)\citenamefont
  {Hausmann}, \citenamefont {Shields}, \citenamefont {Quan}, \citenamefont
  {Chu}, \citenamefont {de~Leon}, \citenamefont {Evans}, \citenamefont {Burek},
  \citenamefont {Zibrov}, \citenamefont {Markham}, \citenamefont {Twitchen}
  \emph {et~al.}}]{hausmann2013coupling}%
  \BibitemOpen
  \bibfield  {author} {\bibinfo {author} {\bibfnamefont {B.~J.}\ \bibnamefont
  {Hausmann}}, \bibinfo {author} {\bibfnamefont {B.~J.}\ \bibnamefont
  {Shields}}, \bibinfo {author} {\bibfnamefont {Q.}~\bibnamefont {Quan}},
  \bibinfo {author} {\bibfnamefont {Y.}~\bibnamefont {Chu}}, \bibinfo {author}
  {\bibfnamefont {N.~P.}\ \bibnamefont {de~Leon}}, \bibinfo {author}
  {\bibfnamefont {R.}~\bibnamefont {Evans}}, \bibinfo {author} {\bibfnamefont
  {M.~J.}\ \bibnamefont {Burek}}, \bibinfo {author} {\bibfnamefont {A.~S.}\
  \bibnamefont {Zibrov}}, \bibinfo {author} {\bibfnamefont {M.}~\bibnamefont
  {Markham}}, \bibinfo {author} {\bibfnamefont {D.~J.}\ \bibnamefont
  {Twitchen}},  \emph {et~al.},\ }\href@noop {} {\bibfield  {journal} {\bibinfo
   {journal} {Nano Lett.}\ }\textbf {\bibinfo {volume} {13}},\ \bibinfo {pages}
  {5791} (\bibinfo {year} {2013})}\BibitemShut {NoStop}%
\bibitem [{\citenamefont {Riedrich-M{\"o}ller}\ \emph
  {et~al.}(2012)\citenamefont {Riedrich-M{\"o}ller}, \citenamefont {Kipfstuhl},
  \citenamefont {Hepp}, \citenamefont {Neu}, \citenamefont {Pauly},
  \citenamefont {M{\"u}cklich}, \citenamefont {Baur}, \citenamefont {Wandt},
  \citenamefont {Wolff}, \citenamefont {Fischer} \emph
  {et~al.}}]{riedrich2012one}%
  \BibitemOpen
  \bibfield  {author} {\bibinfo {author} {\bibfnamefont {J.}~\bibnamefont
  {Riedrich-M{\"o}ller}}, \bibinfo {author} {\bibfnamefont {L.}~\bibnamefont
  {Kipfstuhl}}, \bibinfo {author} {\bibfnamefont {C.}~\bibnamefont {Hepp}},
  \bibinfo {author} {\bibfnamefont {E.}~\bibnamefont {Neu}}, \bibinfo {author}
  {\bibfnamefont {C.}~\bibnamefont {Pauly}}, \bibinfo {author} {\bibfnamefont
  {F.}~\bibnamefont {M{\"u}cklich}}, \bibinfo {author} {\bibfnamefont
  {A.}~\bibnamefont {Baur}}, \bibinfo {author} {\bibfnamefont {M.}~\bibnamefont
  {Wandt}}, \bibinfo {author} {\bibfnamefont {S.}~\bibnamefont {Wolff}},
  \bibinfo {author} {\bibfnamefont {M.}~\bibnamefont {Fischer}},  \emph
  {et~al.},\ }\href@noop {} {\bibfield  {journal} {\bibinfo  {journal} {Nat.
  Nanotechnol.}\ }\textbf {\bibinfo {volume} {7}},\ \bibinfo {pages} {69}
  (\bibinfo {year} {2012})}\BibitemShut {NoStop}%
\bibitem [{\citenamefont {Lee}\ \emph {et~al.}(2012)\citenamefont {Lee},
  \citenamefont {Aharonovich}, \citenamefont {Magyar}, \citenamefont {Rol},\
  and\ \citenamefont {Hu}}]{lee2012coupling}%
  \BibitemOpen
  \bibfield  {author} {\bibinfo {author} {\bibfnamefont {J.~C.}\ \bibnamefont
  {Lee}}, \bibinfo {author} {\bibfnamefont {I.}~\bibnamefont {Aharonovich}},
  \bibinfo {author} {\bibfnamefont {A.~P.}\ \bibnamefont {Magyar}}, \bibinfo
  {author} {\bibfnamefont {F.}~\bibnamefont {Rol}}, \ and\ \bibinfo {author}
  {\bibfnamefont {E.~L.}\ \bibnamefont {Hu}},\ }\href@noop {} {\bibfield
  {journal} {\bibinfo  {journal} {Opt. Express}\ }\textbf {\bibinfo {volume}
  {20}},\ \bibinfo {pages} {8891} (\bibinfo {year} {2012})}\BibitemShut
  {NoStop}%
\bibitem [{\citenamefont {Kuhn}\ \emph {et~al.}(2002)\citenamefont {Kuhn},
  \citenamefont {Hennrich},\ and\ \citenamefont
  {Rempe}}]{kuhn2002deterministic}%
  \BibitemOpen
  \bibfield  {author} {\bibinfo {author} {\bibfnamefont {A.}~\bibnamefont
  {Kuhn}}, \bibinfo {author} {\bibfnamefont {M.}~\bibnamefont {Hennrich}}, \
  and\ \bibinfo {author} {\bibfnamefont {G.}~\bibnamefont {Rempe}},\
  }\href@noop {} {\bibfield  {journal} {\bibinfo  {journal} {Phys. Rev. Lett.}\
  }\textbf {\bibinfo {volume} {89}},\ \bibinfo {pages} {067901} (\bibinfo
  {year} {2002})}\BibitemShut {NoStop}%
\bibitem [{\citenamefont {Tiecke}\ \emph {et~al.}(2014)\citenamefont {Tiecke},
  \citenamefont {Thompson}, \citenamefont {de~Leon}, \citenamefont {Liu},
  \citenamefont {Vuleti{\'c}},\ and\ \citenamefont
  {Lukin}}]{tiecke2014nanophotonic}%
  \BibitemOpen
  \bibfield  {author} {\bibinfo {author} {\bibfnamefont {T.}~\bibnamefont
  {Tiecke}}, \bibinfo {author} {\bibfnamefont {J.}~\bibnamefont {Thompson}},
  \bibinfo {author} {\bibfnamefont {N.}~\bibnamefont {de~Leon}}, \bibinfo
  {author} {\bibfnamefont {L.}~\bibnamefont {Liu}}, \bibinfo {author}
  {\bibfnamefont {V.}~\bibnamefont {Vuleti{\'c}}}, \ and\ \bibinfo {author}
  {\bibfnamefont {M.}~\bibnamefont {Lukin}},\ }\href@noop {} {\bibfield
  {journal} {\bibinfo  {journal} {Nature (London)}\ }\textbf {\bibinfo {volume}
  {508}},\ \bibinfo {pages} {241} (\bibinfo {year} {2014})}\BibitemShut
  {NoStop}%
\bibitem [{\citenamefont {Kok}\ \emph {et~al.}(2007)\citenamefont {Kok},
  \citenamefont {Munro}, \citenamefont {Nemoto}, \citenamefont {Ralph},
  \citenamefont {Dowling},\ and\ \citenamefont {Milburn}}]{kok2007linear}%
  \BibitemOpen
  \bibfield  {author} {\bibinfo {author} {\bibfnamefont {P.}~\bibnamefont
  {Kok}}, \bibinfo {author} {\bibfnamefont {W.~J.}\ \bibnamefont {Munro}},
  \bibinfo {author} {\bibfnamefont {K.}~\bibnamefont {Nemoto}}, \bibinfo
  {author} {\bibfnamefont {T.~C.}\ \bibnamefont {Ralph}}, \bibinfo {author}
  {\bibfnamefont {J.~P.}\ \bibnamefont {Dowling}}, \ and\ \bibinfo {author}
  {\bibfnamefont {G.}~\bibnamefont {Milburn}},\ }\href@noop {} {\bibfield
  {journal} {\bibinfo  {journal} {Rev. Mod. Phys.}\ }\textbf {\bibinfo {volume}
  {79}},\ \bibinfo {pages} {135} (\bibinfo {year} {2007})}\BibitemShut
  {NoStop}%
\bibitem [{\citenamefont {Childress}\ \emph {et~al.}(2005)\citenamefont
  {Childress}, \citenamefont {Taylor}, \citenamefont {S{\o}rensen},\ and\
  \citenamefont {Lukin}}]{childress2005fault}%
  \BibitemOpen
  \bibfield  {author} {\bibinfo {author} {\bibfnamefont {L.}~\bibnamefont
  {Childress}}, \bibinfo {author} {\bibfnamefont {J.~M.}\ \bibnamefont
  {Taylor}}, \bibinfo {author} {\bibfnamefont {A.~S.}\ \bibnamefont
  {S{\o}rensen}}, \ and\ \bibinfo {author} {\bibfnamefont {M.~D.}\ \bibnamefont
  {Lukin}},\ }\href@noop {} {\bibfield  {journal} {\bibinfo  {journal} {Phys.
  Rev. A}\ }\textbf {\bibinfo {volume} {72}},\ \bibinfo {pages} {052330}
  (\bibinfo {year} {2005})}\BibitemShut {NoStop}%
\bibitem [{\citenamefont {Edmonds}\ \emph {et~al.}(2008)\citenamefont
  {Edmonds}, \citenamefont {Newton}, \citenamefont {Martineau}, \citenamefont
  {Twitchen},\ and\ \citenamefont {Williams}}]{edmonds2008electron}%
  \BibitemOpen
  \bibfield  {author} {\bibinfo {author} {\bibfnamefont {A.~M.}\ \bibnamefont
  {Edmonds}}, \bibinfo {author} {\bibfnamefont {M.~E.}\ \bibnamefont {Newton}},
  \bibinfo {author} {\bibfnamefont {P.~M.}\ \bibnamefont {Martineau}}, \bibinfo
  {author} {\bibfnamefont {D.~J.}\ \bibnamefont {Twitchen}}, \ and\ \bibinfo
  {author} {\bibfnamefont {S.~D.}\ \bibnamefont {Williams}},\ }\href@noop {}
  {\bibfield  {journal} {\bibinfo  {journal} {Phys. Rev. B}\ }\textbf {\bibinfo
  {volume} {77}},\ \bibinfo {pages} {245205} (\bibinfo {year}
  {2008})}\BibitemShut {NoStop}%
\bibitem [{\citenamefont {Weber}\ \emph {et~al.}(2010)\citenamefont {Weber},
  \citenamefont {Koehl}, \citenamefont {Varley}, \citenamefont {Janotti},
  \citenamefont {Buckley}, \citenamefont {Van~de Walle},\ and\ \citenamefont
  {Awschalom}}]{weber2010quantum}%
  \BibitemOpen
  \bibfield  {author} {\bibinfo {author} {\bibfnamefont {J.}~\bibnamefont
  {Weber}}, \bibinfo {author} {\bibfnamefont {W.}~\bibnamefont {Koehl}},
  \bibinfo {author} {\bibfnamefont {J.}~\bibnamefont {Varley}}, \bibinfo
  {author} {\bibfnamefont {A.}~\bibnamefont {Janotti}}, \bibinfo {author}
  {\bibfnamefont {B.}~\bibnamefont {Buckley}}, \bibinfo {author} {\bibfnamefont
  {C.}~\bibnamefont {Van~de Walle}}, \ and\ \bibinfo {author} {\bibfnamefont
  {D.~D.}\ \bibnamefont {Awschalom}},\ }\href@noop {} {\bibfield  {journal}
  {\bibinfo  {journal} {Proc. Natl. Acad. Sci. U.S.A.}\ }\textbf {\bibinfo
  {volume} {107}},\ \bibinfo {pages} {8513} (\bibinfo {year}
  {2010})}\BibitemShut {NoStop}%
\bibitem [{\citenamefont {Lettow}\ \emph {et~al.}(2010)\citenamefont {Lettow},
  \citenamefont {Rezus}, \citenamefont {Renn}, \citenamefont {Zumofen},
  \citenamefont {Ikonen}, \citenamefont {G{\"o}tzinger},\ and\ \citenamefont
  {Sandoghdar}}]{lettow2010quantum}%
  \BibitemOpen
  \bibfield  {author} {\bibinfo {author} {\bibfnamefont {R.}~\bibnamefont
  {Lettow}}, \bibinfo {author} {\bibfnamefont {Y.}~\bibnamefont {Rezus}},
  \bibinfo {author} {\bibfnamefont {A.}~\bibnamefont {Renn}}, \bibinfo {author}
  {\bibfnamefont {G.}~\bibnamefont {Zumofen}}, \bibinfo {author} {\bibfnamefont
  {E.}~\bibnamefont {Ikonen}}, \bibinfo {author} {\bibfnamefont
  {S.}~\bibnamefont {G{\"o}tzinger}}, \ and\ \bibinfo {author} {\bibfnamefont
  {V.}~\bibnamefont {Sandoghdar}},\ }\href@noop {} {\bibfield  {journal}
  {\bibinfo  {journal} {Phys. Rev. Lett.}\ }\textbf {\bibinfo {volume} {104}},\
  \bibinfo {pages} {123605} (\bibinfo {year} {2010})}\BibitemShut {NoStop}%
\bibitem [{\citenamefont {Patel}\ \emph {et~al.}(2010)\citenamefont {Patel},
  \citenamefont {Bennett}, \citenamefont {Farrer}, \citenamefont {Nicoll},
  \citenamefont {Ritchie},\ and\ \citenamefont {Shields}}]{patel2010two}%
  \BibitemOpen
  \bibfield  {author} {\bibinfo {author} {\bibfnamefont {R.~B.}\ \bibnamefont
  {Patel}}, \bibinfo {author} {\bibfnamefont {A.~J.}\ \bibnamefont {Bennett}},
  \bibinfo {author} {\bibfnamefont {I.}~\bibnamefont {Farrer}}, \bibinfo
  {author} {\bibfnamefont {C.~A.}\ \bibnamefont {Nicoll}}, \bibinfo {author}
  {\bibfnamefont {D.~A.}\ \bibnamefont {Ritchie}}, \ and\ \bibinfo {author}
  {\bibfnamefont {A.~J.}\ \bibnamefont {Shields}},\ }\href@noop {} {\bibfield
  {journal} {\bibinfo  {journal} {Nature Photon.}\ }\textbf {\bibinfo {volume}
  {4}},\ \bibinfo {pages} {632} (\bibinfo {year} {2010})}\BibitemShut {NoStop}%
\bibitem [{\citenamefont {Flagg}\ \emph {et~al.}(2010)\citenamefont {Flagg},
  \citenamefont {Muller}, \citenamefont {Polyakov}, \citenamefont {Ling},
  \citenamefont {Migdall},\ and\ \citenamefont
  {Solomon}}]{flagg2010interference}%
  \BibitemOpen
  \bibfield  {author} {\bibinfo {author} {\bibfnamefont {E.~B.}\ \bibnamefont
  {Flagg}}, \bibinfo {author} {\bibfnamefont {A.}~\bibnamefont {Muller}},
  \bibinfo {author} {\bibfnamefont {S.~V.}\ \bibnamefont {Polyakov}}, \bibinfo
  {author} {\bibfnamefont {A.}~\bibnamefont {Ling}}, \bibinfo {author}
  {\bibfnamefont {A.}~\bibnamefont {Migdall}}, \ and\ \bibinfo {author}
  {\bibfnamefont {G.~S.}\ \bibnamefont {Solomon}},\ }\href {\doibase
  10.1103/PhysRevLett.104.137401} {\bibfield  {journal} {\bibinfo  {journal}
  {Phys. Rev. Lett.}\ }\textbf {\bibinfo {volume} {104}},\ \bibinfo {pages}
  {137401} (\bibinfo {year} {2010})}\BibitemShut {NoStop}%
\bibitem [{\citenamefont {Neu}\ \emph {et~al.}(2012)\citenamefont {Neu},
  \citenamefont {Agio},\ and\ \citenamefont {Becher}}]{neu2012photophysics}%
  \BibitemOpen
  \bibfield  {author} {\bibinfo {author} {\bibfnamefont {E.}~\bibnamefont
  {Neu}}, \bibinfo {author} {\bibfnamefont {M.}~\bibnamefont {Agio}}, \ and\
  \bibinfo {author} {\bibfnamefont {C.}~\bibnamefont {Becher}},\ }\href@noop {}
  {\bibfield  {journal} {\bibinfo  {journal} {Opt. Express}\ }\textbf {\bibinfo
  {volume} {20}},\ \bibinfo {pages} {19956} (\bibinfo {year}
  {2012})}\BibitemShut {NoStop}%
\end{thebibliography}%
\end{document}